\documentclass{evn2002}
\begin{document}
   \title{Searching for low mass objects around nearby dMe radio stars}

\author{J. C. Guirado\inst{1} \and E. Ros\inst{2} \and 
D. L. Jones\inst{3} \and W. Alef\inst{2} \and J. M. Marcaide\inst{1}
 \and R. A. Preston\inst{3} }

   \institute{Departamento de Astronom\'{\i}a, Universidad de Valencia, 
              46100 Burjassot, Valencia, Spain \and
              Max-Planck-Institut f\"ur Radioastronomie, Auf dem H\"ugel 69, 
              53121 Bonn, Germany \and
              Jet Propulsion Laboratory, California Institute of Technology,
              Pasadena, California 91109, USA }

   \abstract{
   
Nearby M-dwarfs are best suited for searches of low mass companions.
VLBI phase-referencing observations with sensitive telescopes are able to
detect radio star flux-densities of tenths of mJy as well as to position 
the star on the sky with submilliarcsecond precision. We have initiated 
a long-term observational program, using EVN telescopes in combination with 
NASA DSN dishes, to revisit the kinematics of nearby, single M dwarfs. The 
precision of the astrometry allows us to search for possible companions with 
masses down to 1 Jupiter mass. In this contribution we report preliminary 
results of the first observation epochs, in which we could detect some of the 
radio stars included in our program.}
% Two of them, EV\,Lacertae and Wolf\,47, 
%have been detected for the first time with VLBI. }

\titlerunning{Searching for low mass objects around nearby radio stars}
\authorrunning{J.C. Guirado et al.}

   \maketitle 
%
%________________________________________________________________

\section{Introduction}

\noindent
In the last years, optical spectroscopy techniques 
have shown to be extraordinarily effective in detecting very low mass objects 
around nearby stars. However, astrometric techniques still offer two extra 
advantages: first, the determination of the mass of the
companion without coupling with the orbit inclination, and second,
the detection of companions with longer
periods and larger distances from the host star than those typical of the
orbits detected by spectroscopy (that is, orbits similar to
the ones known in our solar system). In particular, submilliarcsecond VLBI 
astrometry has demonstrated its capabilities for more than two decades. 
%Furthermore, VLBI astrometry, using the phase-referencing technique, has 
%already succeeded in the detection of low-mass companions orbiting nearby 
%weakly-emitting radio stars (Guirado et al. 1997). 

VLBI phase-referencing allows the detection of weak radio sources, with 
flux densities smaller than the sensitivity limit imposed by the fluctuations 
of the atmosphere and the frequency standard at each VLBI site. A reference 
for the VLBI phase must be established by interleaving observations of an 
angularly nearby strong extragalactic source. This cyclic observing scheme 
effectively allows to increase the coherent integration time on the target 
(weak) source from minutes to hours, with the corresponding improvement of 
the sensitivity. In addition, since the main observable used is the
interferometric phase, this technique provides high-precision relative 
position information of the two observed sources.

Among the stellar objects potentially detectable with VLBI phase-referencing, 
nearby, single radio-emitting stars are best suited for astrometric searches 
for low-mass stellar companions. In particular, nearby M-dwarfs are very good 
candidates: they are close to the Sun ($\sim$5 pc), and they 
have very low inertia (M$\sim$0.3\,M$_\odot$, for a main-sequence M5 dwarf) 
which, according to the Kepler's third law, implies a large reflex motion for
a given orbiting low-mass companion. In addition to this, many M dwarfs 
exhibit non-thermal, compact radio emission consisting of a slowly varying 
quiescent flux that may increase by an order of magnitude when the star is 
flaring.

We have initiated a long-term VLBI astrometric program to observe a sample of 
nearby dMe stars. The goal of this program is to refine the 
kinematics of nearby radio stars to uncover possible reflex motions resulting 
from the gravitational interaction of unseen low mass companions.
From the very active, well-known flare M dwarfs, we 
have selected those placed 
around 10pc from the Sun, detected previously at least with the VLA at a flux 
density of 1 mJy or more (see Table 1). Some of 
them have already been detected with VLBI arrays: AD\,Leo, YZ\,CMi  
(Pestalozzi et al. 2000 and references therein), EQ\,Peg\,B (Benz et al 1995), 
Wolf\,630\,A, and EV\,Lac (Phillips et al. 1989). 
In our sample there are both single stars (EV\,Lac, YZ\,CMi, AD\,Leo, and Wolf\,47), 
already included in ongoing
planetary searches in the visible and infrared, and wide-separation binaries, with 
the partner star placed a few arcseconds away (3" and 5" for DO\,Cep and EQ\,Peg\,B, 
respectively). In the latter case, the gravitational interaction produces a long-term 
orbital motion that should be disentangled easily from the possible reflex wobble 
corresponding to another companion. In this contribution, we report the first 
detections of radio stars included in our sample. Despite of the preliminary astrometric 
analysis, these detections show the feasibility of our program to detect companions 
down to 1 Jupiter mass.

\begin{table*} 
\caption{Radio stars included in our program}
\begin{center}
\begin{tabular}{lccc}\hline 
              & {Distance}  &  {Quiescent $^\mathrm{a}$} & {Reference $^\mathrm{b}$} \\
              & {(pc)}      & {emission (mJy)}     &  {source}            \\ \hline
Wolf\,47      &  9.3            &   $<$0.3 - 7.1           & 0059+581 (3.6\degr)  \\ 
YZ\,CMi       &  6.1            &   0.6 - 1.5              & 0736+017 (2.4\degr)  \\ 
AD\,Leo       &  4.9            &   0.2 - 2.1              & 1022+194 (1.4\degr)  \\ 
DO\,Cep       &  4.0            &   $<$0.4 - 5.5           & 2250+555 (3.1\degr)  \\ 
EV\,Lac       &  5.1            &   $<$0.3 - 4.0           & 2253+417 (2.8\degr)  \\ 
Wolf\,630\,A  &  6.2            &   $<$2.0 - 4.0           & 1741$-$038 (5.5\degr)  \\ 
EQ\,Peg\,B    &  6.6            &   1.1 - 5.5              & J2341+19 (2.2\degr)  \\ \hline
\end{tabular}
\end{center}
\begin{footnotesize}
%\vspace{0.2cm}
\noindent
$^\mathrm{a}$: Slowly varying quiescent, non-flare emission 
at cm-wavelengths (taken from Caillault et al. 1988; Fomalont \& Sanders 1989; Phillips et al. 1989; 
White et al. 1989; Benz \& Alef 1991; Benz et al. 1995; Leto et al. 2000). 
$^\mathrm{b}$: Radio sources selected to act 
as reference sources in the 
phase-reference mapping analysis. Separation from the 
corresponding radio stars in parentheses.
% A second reference source may 
%be used to avoid ambiguity in the position determination and drive 
%systematic errors lower.\\
\end{footnotesize}
\end{table*}

%__________________________________________________________________

\section{Observations and Data Reduction}

We made VLBI phase-referenced observations of dMe stars 
at 8.4 GHz at multiple epochs (see Table 2) with EVN antennas along with the 
70m DSN dishes in 
Madrid (DSS63) and Goldstone (DSS14). For each experiment, we interleaved 
observations of a strong background radio source. The data were correlated with 
the Mark III/IV correlator at the MPIfR in Bonn. For each epoch and radio star, 
we derived a priori (correlation) coordinates
from the data given in the Hipparcos catalogue (ESA 1997). We corrected
these coordinates for proper motion and trigonometric parallax effects 
from a reference epoch (1991.25 for the Hipparcos catalogue) to 
the epoch of our observations. For each experiment, we performed a 
phase-calibration analysis within AIPS 
(Beasley \& Conway 1995). The resulting referenced phases of the radio stars 
show the corrections of the a priori relative coordinates. We determined 
these corrections by inspection of the phase-reference map of the radio star, 
constructed by Fourier inversion of its (referenced) visibilities.

\begin{table*} 
%%%\begin{small}
\caption{Radio star VLBI observations}
\begin{center}
\noindent
\begin{tabular}{rlc}\hline 
Obs. Date               & Stations           & Stars observed         \\ \hline
28-Oct-1996 (1996.82)   &  Bonn, DSS65, Onsala, Noto &  EQ\,Peg\,B  \\ 
21-Aug-1999 (1999.64)   &  Bonn, DSS63, DSS14        &  Wolf\,47, DO\,Cep, EV\,Lac  \\ 
6-Oct-1999  (1999.76)   &  Bonn, DSS63               &  Wolf\,47, EV\,Lac         \\ 
11-Dec-1999 (1999.94)   &  Bonn, DSS63               &  AD\,Leo, DO\,Cep, EQ\,Peg\,B  \\ 
8-Jan-2000  (2000.02)   &  Bonn, DSS63               &  AD\,Leo \\ 
4-Jul-2000  (2000.50)   &  Bonn, DSS63               &  DO\,Cep, EV\,Lac      \\
5-Feb-2001  (2001.10)   &  Bonn, DSS63, DSS14        &  DO\,Cep, EV\,Lac      \\ \hline
\end{tabular}
\end{center}
%\noindent
%{\bf (a)} B, Effelsberg; M, DSS63; D, DSS14\\
%{\bf (b)} MarkIV experiment\\ 
\end{table*} 

The sensitivity of our array is largely dominated by the performance of the 
baseline DSS63-Bonn. The use of short arrays for radio star astrometry was  
proved to be successful: with a similar array (DSS43--Hobart) the 
kinematics of some southern radio stars was monitored to 
the the precision of a tenth of milliarcsecond. The precision 
and consistency of those data allowed even the detection of a previously unseen 
low-mass object around the star AB\,Doradus (Guirado et al. 1997). 
For comparison, the baseline DSS63--Bonn provides similar precision in the 
astrometry ($\sim$0.3\,mas in the determination of the relative positions) 
and far more sensitivity (threshold sensitivity of 0.4 mJy at 8.4\,GHz, 
corresponding to 7$\sigma$ for 5 hours integration time and 56\,MHz bandwidth).

%-------------------------------------------------------------

   \begin{figure}
   \centering
   \vspace{250pt}
%   \special{psfile=eqpeg2.ps hscale=50 vscale=50 hoffset=-75 voffset=-20}
   \includegraphics{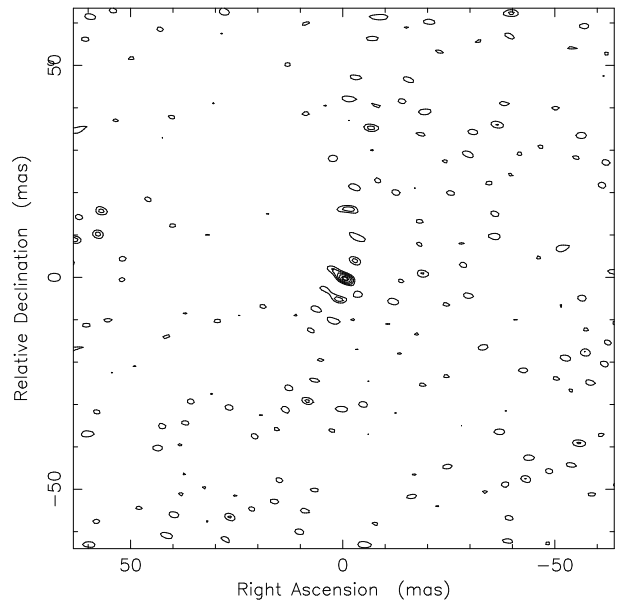}
   \caption{Phase-referenced (dirty) map of EQ\,Peg\,B. The offset of the 
            brightness maximum from the map center corresponds to the error
            in the a priori coordinates of EQ\,Peg\,B relative to 2337+220. 
            Contours are 30, 45, 60, 75, 90, and 95\% of the peak of brightness. 
            The restoring beam is an elliptical Gaussian of 3.8$\times$2.3\,mas 
	    (P.A. 86\degr). }
            \label{fig:eqpeg}
    \end{figure}

%-------------------------------------------------------------

%-------------------------------------------------------------

   \begin{figure}
   \centering
   \vspace{250pt}
%   \special{psfile=evlac2.ps hscale=50 vscale=50 hoffset=-75 voffset=-20}
   \includegraphics{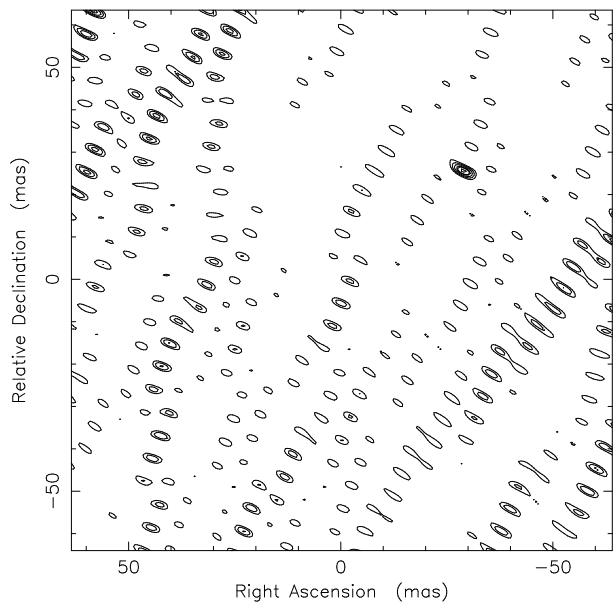}
   \caption{Phase-referenced (dirty) map of EV\,Lac. The offset of the 
            brightness maximum from the map center corresponds to the error 
            in the a priori coordinates of EV\,Lac relative to 2253+417. 
            Contours are 30, 45, 60, 75, 90, and 95\% of the peak of brightness. 
            The restoring beam is an elliptical Gaussian of 5.6$\times$2.0\,mas
	    (PA 69\degr). }
            \label{fig:evlac}
    \end{figure}

%-------------------------------------------------------------

%-------------------------------------------------------------

   \begin{figure}
   \centering
   \vspace{250pt}
%   \special{psfile=wolf472.ps hscale=50 vscale=50 hoffset=-75 voffset=-20}
   \includegraphics{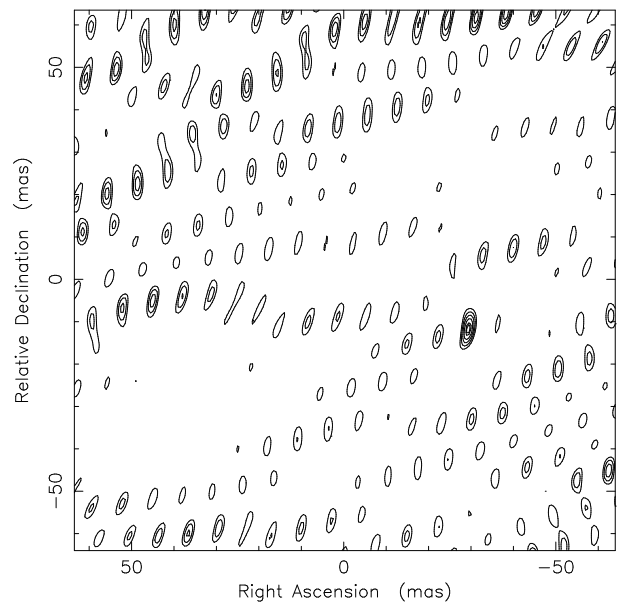}
   \caption{Phase-referenced (dirty) map of Wolf\,47. The offset of the 
            brightness maximum from the map center corresponds to the error
            in the a priori coordinates of Wolf\,47 relative to 0059+581. 
            Contours are 30, 45, 60, 75, 90, and 95\% of the peak of brightness. 
            The restoring beam is an elliptical Gaussian of 6.3$\times$2.4\,mas
	    (PA -11\degr). }
            \label{fig:wolf47}
    \end{figure}

%-------------------------------------------------------------

\section{Results and Discussion}

We have analyzed the first three epochs shown in Table 2. We did detect 
EQ\,Peg\,B at epoch 1996.82, and EV\,Lac and Wolf\,47 at epoch 
1999.76.  We failed to detect any of the stars observed at epoch 
1999.64. In what follows, we give for each source some notes: 

\subsection{EQ\,Peg\,B} This star is a dM6e main-sequence star at a distance 
of 6.6 pc (see Table 1), actually part of a wide-separation 
binary system with the partner star placed at $\sim$5". EQ\,Peg\,B has 
been detected at 18cm with intercontinental baselines (Benz et al. 1995). 
We have detected this star at our first epoch at the level of 2.8 mJy, that is 
more than 10 times the theoretical root-mean-square (rms) noise level. The phase-referenced 
map of EQ\,Peg\,B is displayed in Fig. 1; for each coordinate, the peak
of brightness is shifted less than 1\,milliarcsecond (mas) with respect to 
a priori position.

\subsection{EV\,Lac} This single star is a well-known dM4.5e radio star at a 
distance of 5.1 pc (see Table 1). EV\,Lac is a traditional target in planetary 
searches carried out through radial velocity measurements, infrared 
observations, or optical astrometry. This object is a usual target in stellar 
radio astronomy: it has been detected several times with the VLA 
(White et al. 1989, Leto et al. 2000), and an early detection with VLBI 
arrays (actually using the Bonn-DSS63 baseline) was reported by 
Phillips et al. (1989), with a flux density of 2-4 mJy. 
We have detected EV\,Lac in at our third epoch (1999.76) at the level of 3\,mJy, 
that is $\sim$30 times the theoretical rms noise level. The phase-referenced 
map of EV\,Lac is displayed in Fig. 2; the peak of brightness is shifted 
$-$29\,mas in right ascension and 25\,mas in declination. 

\subsection{Wolf\,47} This star is a dM5e radio star at a 
distance of 9.3 pc (see Table 1). Wolf\,47 
has been detected several times with the VLA at 8.4\,GHz 
(Hewitt et al. 1989) and at longer wavelengths
(White et al. 1989). Although it was proposed as a promising 
VLBI target (Hewitt et al. 1989), there is not any reported 
attempt to observe this radio star with VLBI. 
Hence, our detection of Wolf\,47, with a flux density of 2.2\,mJy, 
$\sim$20 times the theoretical rms noise level, happens to be the 
first one with VLBI. The phase-referenced 
map of EV\,Lac is displayed in Fig. 3; the peak of brightness is shifted 
$-$30\,mas in right ascension and $-$11\,mas in declination.

The phase-referenced maps shown in Figs. 1-3 were derived from 
the Fourier inversion of the referenced visibilities, which relied 
solely on the MkIII correlator astrometric model.
% (which in turn corresponds to that included in the software CALC 8.0). 
Although the astrometric analysis 
is still underway, the large correction from the 
a priori position of some of the detected radio stars can be justified 
by the large standard deviation of the a priori positions (standard 
deviation dominated by the contribution of the error in the proper motion measurement, 
which scales by the $\sim$8-years time interval between the Hipparcos reference 
date and the epoch of our observations), and to a lesser extent, to 
the lack of atmospheric correction in our preliminary astrometric 
analysis. These latter effects are partially canceled by the 
proximity of the background radio source to the radio star on the sky, 
and they are unlikely to contribute more than a few milliarcseconds.

Finally, a word on the expected precision of our astrometric 
program. We expect that fluctuating error will dominate the standard 
deviation of the relative position of the radio star. Among these
errors, instabilities of the star's surface are most important since 
they may change the reference point selected for the astrometry randomly
from epoch to epoch. On this respect we notice that 
i) the size of the photosphere of the proposed stars ranges from 0.2 to 0.8 mas
(White et al. 1989; Delfosse et al. 1998) about the order of magnitude of the 
expected precision of our determination; and ii) the motion of the hot spots
may not have a given trend, therefore it is expected to be averaged out 
from our data after several epochs. Thus, our initial goal to detect orbital 
motions of $\sim$1\,mas amplitude should still be achievable with
our VLBI-based search, enough precision to detect companions with masses
as low as 1 Jupiter mass in less than four years.

%__________________________________________________ One column table
%
%\section{Conclusions}
%
%Everything has a beginning and an end.  That is the end of this sample
%paper.  Have fun by writing your one!
%

\begin{acknowledgements}
We acknowledge the staff of the MPIfR correlator and the observing
radio telescopes.  We are especially grateful to D.A.\ Graham for his help
in preparing the observations.
The European VLBI Network is a joint facility of European, Chinese and
other radio astronomy institutes funded by their national research councils.
This research was supported by the European Commission's IHP Programme
``Access to Large-scale Facilities", under contract No.\ HPRI-CT-1999-00045
We acknowledge the support of the European Comission - Access to Research
Infrastructure action of the Improving Human Potential Programme.
This work has been partially supported by the Spanish MCYT Grant No. 
AYA2001-2147-C02-02. Research at JPL is carried out under contract with 
the National Aeronautics and Space Administration.
\end{acknowledgements}

\end{document}